\documentclass[conference,10pt]{IEEEtran}
\usepackage[utf8]{inputenc}
\usepackage{graphicx}

\usepackage{subfigure}
\usepackage{booktabs}
\usepackage{cleveref}
\usepackage{multirow}
\usepackage{soul}
\usepackage{pgf}

\usepackage{soul}
\usepackage{flushend}
\newtoggle{screen}
\toggletrue{screen}

\usepackage{setspace}
\usepackage[normalem]{ulem}
\usepackage[font=small]{caption}
\usepackage{multirow}
\usepackage{color, colortbl}
\usepackage{footnote}
\usepackage{slashbox,pict2e}
\usepackage{pifont}
\newcommand{\cmark}{\ding{51}}%
\newcommand{\xmark}{\ding{55}}%
\definecolor{cadmiumgreen}{rgb}{0.0, 0.42, 0.24}
\usepackage{fancyhdr}

\title{Is Robust Design-for-Security Robust Enough? \\Attack on Locked Circuits \\with Restricted Scan Chain Access}

\author{Nimisha Limaye, Abhrajit Sengupta, Mohammed Nabeel, and Ozgur Sinanoglu\\
	{\small Tandon School of Engineering, New York University, New York, USA}\\
	{\small Division of Engineering, New York University Abu Dhabi, United Arab Emirates}\\
	{\small \{nsl278, as9397, mtn2, ozgursin\}@nyu.edu}
}

\begin{document}

\maketitle
\renewcommand{\headrulewidth}{0.0pt}
\thispagestyle{fancy}
\pagestyle{fancy}
\cfoot{
\copyright~2019 IEEE.
This is the author's version of the work. It is posted here for your personal use.
	Not for redistribution.\\
	The definitive Version of Record is published in
	Proc. International Conference on Computer-Aided Design (ICCAD), 2019.
}
\begin{abstract}
The security of logic locking has been called into question by various attacks, especially a Boolean satisfiability (SAT) based attack, that exploits scan access in a working chip. Among other techniques, a robust design-for-security (DFS) architecture was presented to restrict any unauthorized scan access, thereby, thwarting the SAT attack (or any other attack that relies on scan access). Nevertheless, in this work, we successfully break this technique 
by recovering the secret key despite the lack of scan access. Our security analysis on a few benchmark circuits protected by the robust DFS architecture demonstrates the effectiveness of our attack; \textcolor{black}{on average $\sim$95\% of the key bits are correctly recovered, and almost 100\% in most cases.} 
To overcome this and other prevailing attacks, we propose a defense by making fundamental changes to the robust DFS technique; the new defense can withstand all logic locking attacks. We observe, on average, lower area overhead ($\sim$1.65\%) than the robust DFS design ($\sim$5.15\%), and similar test coverage ($\sim$99.88\%).

\end{abstract}

\section{Introduction}
The changing landscape of the semiconductor industry has led to many security threats such as intellectual property (IP) piracy~\cite{chipworks}, counterfeiting~\cite{guin2014counterfeit}, and insertion of hardware Trojans~\cite{rostami2014primer}. The prohibitive cost of owning a fabrication facility, which can be up to several billions of dollars, has forced many companies to go fabless over the years. As this trend consolidates, design companies rely on external, \emph{untrusted} foundries for cost-effective access to advanced technology nodes. Due to the lack of any monitoring in this scenario, the trust in the Integrated Circuit (IC) supply chain has been called into question~\cite{rostami2014primer}.

To enable trust in the supply chain, several design-for-security (DFS) techniques were developed such as logic locking, split manufacturing, IC metering, camouflaging, etc.~\cite{rostami2014primer}. Among these techniques, logic locking is perceived as a holistic solution due to its ability to protect against a rogue element at any stage in the supply chain. \textcolor{black}{Logic locking involves the insertion of key gates in the circuit, obfuscating the functionality based on a secret key known only to the designers.} The success of logic locking in protecting a circuit relies on how well the secret key can be protected. Early attempts at breaking logic locking include sensitization attack~\cite{JV_DAC_2012}, test data mining attack~\cite{yasin_DATE_2016}, etc., by retrieving the secret key. However, later a lethal Boolean satisfiability (SAT) based attack completely undermined the security of the existing locking techniques~\cite{Subramanyan_host_2015}.

The post-SAT era saw the research community embarking on two separate directions. On the one hand, several SAT-resilient locking techniques were proposed~\cite{yasin2016sarlock, anti-sat, sfll-fault, delaylock, cyclock}. However, these techniques suffer from various structural vulnerabilities that were eventually exploited to break them~\cite{xu2017novel, removal, sirone2018functional, chakraborty2018timingsat, smt, zhou2017cycsat, yang2019stripped}. Further, all SAT-resilient techniques suffer from low corruptibility/error rate, still allowing an attacker to approximately recover a circuit~\cite{appsat}.

On the other hand, attempts at thwarting SAT attack focused on restricting unauthorized scan access. This branches down to either obfuscating the scan chains~\cite{karmakar2018encrypt, wang2018secure}, or blocking the scan chain access entirely~\cite{guin2018robust,wang2019secure,ahlawat2017preventing}. 
In~\cite{karmakar2018encrypt}, Karmakar et al. proposed obfuscating the stimuli and test patterns by inserting key gates into the scan chain. Nevertheless, it was broken by Alrahis et al. by transforming the locked scan chains into a combinational circuit, and thereby launching the well-known SAT attack against it~\cite{alrahis2019scansat}. 
Recently, a cost-effective robust design-for-security (DFS) architecture was proposed to protect the key of a logic-locked circuit while enabling secure test and debug operations, where a new secure scan cell design, called \emph{secure cell} (SC), was introduced~\cite{guin2018robust}.\footnote{Throughout this paper, we will refer to this defense architecture as DFS.} The logic locking key is held in the SCs securely and its leakage is prevented by \emph{blocking the scan read-out} operations upon a switch from functional to test mode, thereby, thwarting all the attacks that require scan access. 
Further, they also block scan read-out in the functional mode, restricting chip response to be observed only at the primary outputs.

\textbf{Contribution.}
The robust DFS architecture has tremendous potential for thwarting all the existing logic locking attacks, as it blocks scan read-out access.
In this work, we propose a shift-and-leak attack to \emph{break} a circuit secured through this important defense. The attack does not require access to scan read-out; it rather observes the chip responses through chip pin-outs and is applicable even in the restricted scan access setting. Thus, our proposed attack can be contrasted from all the existing attacks which cannot be launched against the DFS architecture. 
The contributions of this work are as follows.

\begin{itemize}
    \item We propose a new shift-and-leak attack that first judiciously moves the circuit from the capture state (which protects the key as per the design of DFS architecture) into a key-leaking state by leveraging shift operations. The attack framework utilizes synthesis and automatic test pattern generation (ATPG) tools to leak key bits one at a time.
    \item We propose a pre-processing step to boost our shift-and-leak attack. This involves launching a SAT
    attack in a \emph{restricted} scan access setting. 
    \item Our experimental results show that \textcolor{black}{on average, our attack can retrieve almost all the key bits ($\sim$95\%).} 
    \item We propose a countermeasure, \textcolor{black}{mode switch shift disable (MSSD),} to thwart our proposed attack as well as existing attacks, and give a rigorous security analysis comparing against different attacks and defenses. As shown in Table~\ref{tab:compare}, our proposed defense is resilient against all the mentioned attacks including our proposed shift-and-leak attack. Essentially, the proposed defense technique can be utilized in conjunction with a basic SAT-vulnerable but high-corruptibility logic locking technique to defend against all existing attacks. Note that AppSAT~\cite{appsat} cannot decompose these defenses in the absence of scan access.
    \item We achieve lower area overhead (on average 1.65\%) compared to DFS architecture (on average 5.15\%). Test coverage is observed to be almost the same for both the proposed and DFS architectures.
\end{itemize}

\begin{table}[!tb]
\caption{Comparison between defenses and attacks. \textcolor{cadmiumgreen}{\cmark} means the defense is resilient to the attack, and \textcolor{red}{\xmark} means the defense is vulnerable to the attack. }
\label{tab:compare}
\centering
\setlength{\tabcolsep}{2pt}
\small\addtolength{\tabcolsep}{.0pt}
\begin{tabular}{|c|c|c|c|c|}
\hline
\backslashbox{{Defense}}{{Attack}}
                        & \textbf{(App)SAT}\cite{Subramanyan_host_2015,appsat} & \textbf{Removal}\cite{removal} & \textbf{ScanSAT}\cite{alrahis2019scansat}  & \textbf{Shift$\&$leak}\\ \hline
\textbf{SSTC}~\cite{ahlawat2017preventing,wang2019secure}       	& \color{cadmiumgreen}{\cmark}          & \color{red}{\xmark} & \color{cadmiumgreen}{\cmark}              & \color{cadmiumgreen}{\cmark}  \\ \hline            \textbf{EFF + RLL}~\cite{karmakar2018encrypt}         & \color{cadmiumgreen}{\cmark}   & \color{cadmiumgreen}{\cmark}          & \color{red}{\xmark}              & \color{cadmiumgreen}{\cmark}                      \\ \hline
\textbf{DFS + SLL}~\cite{guin2018robust}       & \color{cadmiumgreen}{\cmark}  & \color{cadmiumgreen}{\cmark}           & \color{cadmiumgreen}{\cmark}              & \color{red}{\xmark}                      \\ \hline
\textbf{MSSD + RLL}   & \color{cadmiumgreen}{\cmark}       & \color{cadmiumgreen}{\cmark}      & \color{cadmiumgreen}{\cmark}              & \color{cadmiumgreen}{\cmark}                     \\ \hline
\end{tabular}
\end{table}
 
\section{Background}

The design-for-security (DFS) architecture, proposed in~\cite{guin2018robust}, has two goals in mind: 1) protecting scan interface of a logic locked circuit from unauthorized access to thwart SAT, sensitization, etc. attacks on logic locking, and 2) yet preserving the full testability and debug of the chip. The architecture ensures no leakage of key bits through the use of secure cells and blocked scan read-out, while supporting structural/manufacturing tests, post-silicon validation and debug, and full in-system test capability. Next, we briefly describe the architecture that is presented in~\cite{guin2018robust}.

\begin{figure}[!tb]
\centering
\subfigure[]{\includegraphics[width=0.3\textwidth]{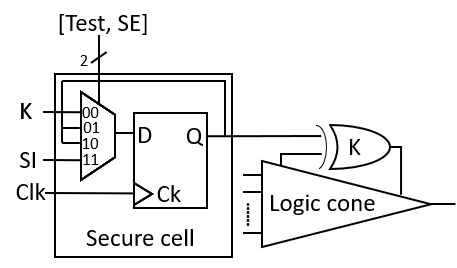}\label{fig:secure_cell}}
\hfill
\subfigure[]{\includegraphics[width=0.4\textwidth]{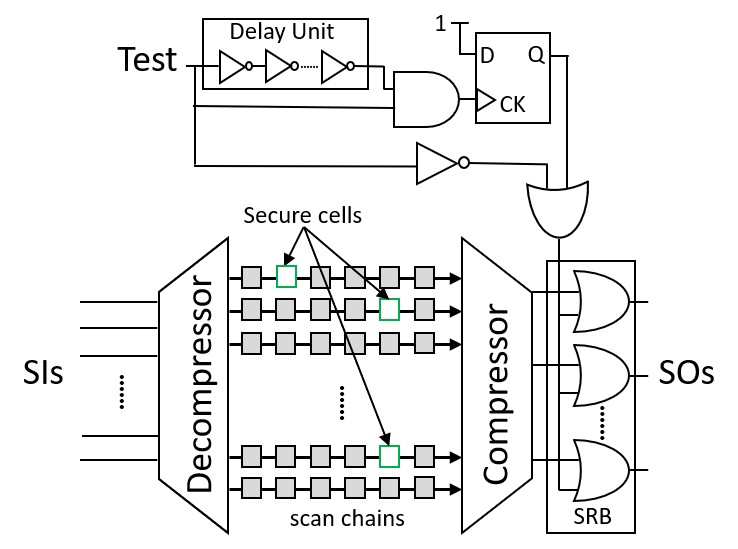}\label{fig:block_scan}}
\hfill
\caption{(a) Secure Cell (SC) design. (b) Architecture with restricted scan access. SRB stands for scan read-out block. Scan read-out is blocked whenever Test signal is low or has a positive transition. Scan load has no restrictions. Source:~\cite{guin2018robust}.}
\end{figure}

\textbf{Architecture.}
The key component is a new scan cell, called a secure cell (SC); one instance of SC is added to the design for each key bit. The SCs are stitched together with regular scan cells (RCs). As illustrated in Fig.~\ref{fig:secure_cell}, the key gate $K$ is driven by the SC holding the correct key bit. SC supports three modes of operation: $M0$, $M1$, and $M2$, determined by two select signals, namely, \emph{Test} and \emph{scan enable (SE)}. The working of the three modes is shown in Table~\ref{tab:modes}. Mode $M0$ denotes the functional mode; the SC captures the correct key bit coming from the tamper-proof memory.
In $M1$, the SC is bypassed, retaining its content, while the RCs shift or capture based on the value of SE. Finally, in mode $M2$, the SC becomes a part of the scan chain along with \textcolor{black}{the} other RCs; the key bits inside the SCs are overwritten by the shifted stimuli (consisting of dummy key bits) and responses, enabling structural tests securely.

Apart from the above three modes, an integral part of this architecture is that the scan read-out is blocked upon a switch from a mode where the key is held in the SCs, into a mode that supports shift operation. More specifically, scan read-out is blocked upon a positive transition on the Test pin or while the Test signal is 0.

The overview of the architecture is presented in Fig.~\ref{fig:block_scan}. SIs and SOs indicate the scan inputs and outputs, respectively. A masking unit with an array of OR gates blocks the scan read-out operations as explained above. Stimulus shift-in through scan inputs can still be performed without any restriction; this capability is preserved in the DFS architecture to support test and debug operations. We later show that our attack exploits this capability. 

\begin{table*}[!tb]
\caption{The three modes of operation for the secure cell.}
\label{tab:modes}

\begin{tabular}{|c|c|c|c|c|}
\hline
\textbf{Test} & \textbf{SE} & \textbf{Mode}       & \textbf{Description}                                                                                                                                                           & \textbf{Testing/Debugging}                                                                                              \\ \hline\hline
0             & 0           & M0                  & \begin{tabular}[c]{@{}c@{}}The chip is in functional mode. \\ The secure cell captures and applies key to the logic.\end{tabular}                                                           & \begin{tabular}[c]{@{}c@{}}Functional testing.\end{tabular} \\ \hline
0             & 1           & M1a & \multirow{2}{*}{\begin{tabular}[c]{@{}c@{}}The secure cell holds its previous value. The rest of the \\ circuit is in shift/functional mode depending on the SE \textcolor{black}{value}.\end{tabular}} & \multirow{2}{*}{\begin{tabular}[c]{@{}c@{}}Shift ($M1$a) to support functional tests. \\Capture ($M1$b) to support structural tests.\end{tabular}}                                                                                               \\ \cline{1-3}
1             & 0           & M1b                    &                                                                                                                                                                                &                                                                                                                         \\ \hline
1             & 1           & M2                  & \begin{tabular}[c]{@{}c@{}}The secure cell becomes a part of the scan chain(s).\end{tabular}                                                                   & \begin{tabular}[c]{@{}c@{}}Shift in/shift out to support structural tests.\end{tabular}                                                                                                                     \\ \hline
\end{tabular}
\end{table*}

\textbf{Test/debug operations.} Structural testing is performed (by untrusted parties) by holding Test signal at 1; shift and capture operations are performed in $M2$ and $M1$b, respectively, with unrestricted scan access, but by loading dummy key bits into SCs. 
Functional testing is performed by first loading the secret key from the memory into SCs in $M0$, and then loading the initial state into the RCs in $M1$a while holding the key value in SCs. The response is then observed at the primary outputs in $M0$.

\begin{figure}[!tb]
    \centering
    \footnotesize
    \includegraphics[width=0.48\textwidth]{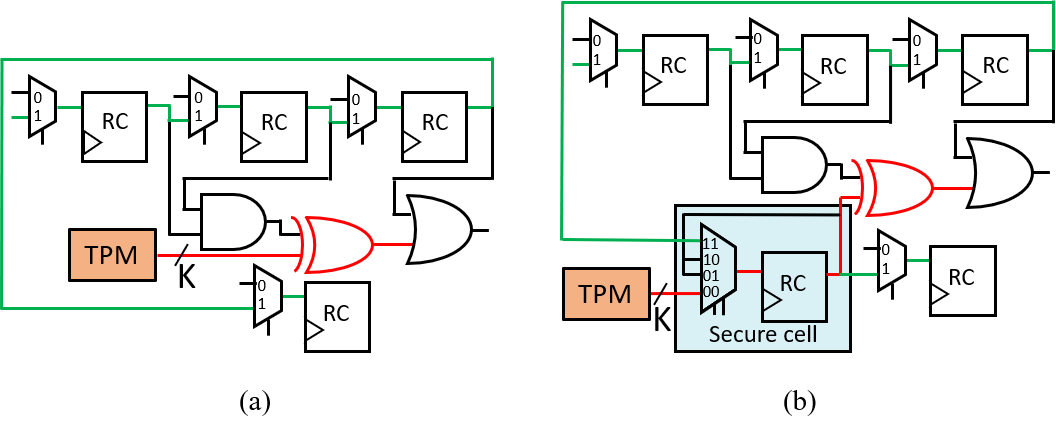}
    \caption{Logic locking applied (a) standalone and (b) in conjunction with DFS. Path shown in red is the key path; path shown in green is the scan path. TPM is the tamper-proof memory where the key is stored.}
    \label{fig:ll_dfs}
\end{figure}

\textbf{Security properties.} In~\cite{guin2018robust}, the authors claim that any attack that aims at leaking the key bits through the scan chains, such as SAT and sensitization attacks, is thwarted. The underlying reason is that any mode transition that involves the leakage of the secret key into \textcolor{black}{the RCs} triggers a scan read-out block operation. Moreover, the authors recommend using strong logic locking (SLL)~\cite{JV_DAC_2012}, so as to prevent leakage of key bits by any \emph{sensitization} attack. This way, {\em they secure the mode that the circuit moves into right after the capture operation}; the key bits maximally interfere with each other in the post-capture state. \textcolor{black}{Figures~\ref{fig:ll_dfs}(a) and~\ref{fig:ll_dfs}(b) show how logic locking can be applied standalone and in conjunction with DFS, respectively.}

By manipulating the state through shift operations and leveraging the fact that primary outputs can (and should) still be observed on the chip pin-outs in the functional mode, our attack can circumvent this defense.

\section{Proposed Attack}
In the DFS architecture, any unauthorized access to the scan chain is restricted by the scan read-out block. This naturally raises the question \emph{is it possible to launch an attack even in the scan read-out block setting?}

In this section, we answer the above question affirmatively, where we present an attack on the robust DFS architecture; our approach is to leak key bits through the primary outputs (POs) of a working chip even with \textcolor{black}{the} restricted scan access.

\textbf{Threat model.} Note that our attack is performed assuming the traditional threat model for logic locking~\cite{Subramanyan_host_2015, JV_DAC_2012}, where:
\begin{enumerate}
    \item The attacker has all the structural information from a \emph{reverse-engineered} locked netlist, including the knowledge of the secure cells, but is missing the correct key bit values, and
    \item The attacker has access to a working chip that embeds the correct key in its secure memory. Note that our attack is carried out with \emph{scan read-out block}, which is a departure from the traditional setting where the attacker has unrestricted access to scan chains.
\end{enumerate}

\subsection{Shift-and-leak attack}
\subsubsection{Idea.} 
As per the architecture, key gates are driven by the secure cells (SCs) containing the key, which are stitched together with regular cells (RCs) into scan chains. The attack first identifies scan cells (mainly RCs but sometimes SCs as well) that can leak information onto a primary output; we call such cells \textcolor{black}{as \textit{leaky cells}} (LCs). The secret key bit in a SC is moved into a LC via shift operations (the shift part of Shift-and-Leak) for as many cycles as the scan distance between the SC and the LC.\footnote{Scan distance is the difference in the position of two scan cells in a scan chain.} The LC is then propagated to one of the primary outputs through the combinational logic (the leak part of Shift-and-Leak). For this, a leak condition needs to hold; the content of the scan chains should fulfill the leak condition during that clock cycle.

\subsubsection{Methodology.} 
The challenge lies in applying the attack in the scan read-out block setting. As the scan read-out is blocked in the $M1$a mode and upon a switch from $M0$ to $M2$, the authors of~\cite{guin2018robust} argue that the content of the SCs cannot be leaked. However, the architecture still allows for the scan load in an unrestricted manner to support test and debug operations. Furthermore, the primary output ports of the chip are still observable.\footnote{As will be explained later in more detail, our attack can bypass the boundary scan cells that normally control and observe the logic that drives the POs; we simply perform a clock-less switch from $M1$a back to $M0$, i.e., the functional mode, where the primary output ports are driven by the combinational logic directly.} The attack is launched as follows.

\begin{enumerate}
    \item Identify LC candidates by extracting the combinational fan-in cones of POs.
    \item Insert a stuck-at-fault at the chosen LC candidate.
    \item Run automatic test pattern generation (ATPG) tool to observe this fault with all SCs set to unknown x's during this process. The ATPG tool returns a test pattern, if it can identify one. This pattern is the leak condition that the scan chain content must meet to leak LC onto a PO. If the ATPG tool cannot identify a test pattern, then rule out this candidate as an LC and target another LC candidate; repeat steps 2 and 3.
    \item Boot the chip in $M0$ to load the secret key in the SCs.
    \item Let $d$ denote the scan distance between the SC and the LC. Now, change the mode to $M1$a and start shifting the $d$-bit reverse-shifted version of the leak condition into the scan chains while SCs hold their values.\footnote{\textcolor{black}{Attacker has access to the Test and SE pins, and hence, switching between modes can easily be carried out.}} 
    \item Post $M1$a, switch to $M2$, and the SCs become part of the scan chain. Now, perform a $d$-bit shift to have the leak condition formed in the scan chains, while the target key bit in the SC gets shifted to LC.
    \item Lastly, clocklessly switch to $M0$ and observe the PO to leak the content of the LC, i.e., the target key bit.
\end{enumerate}

\begin{figure}[!tb]
    \centering
     \includegraphics[width=0.45\textwidth]{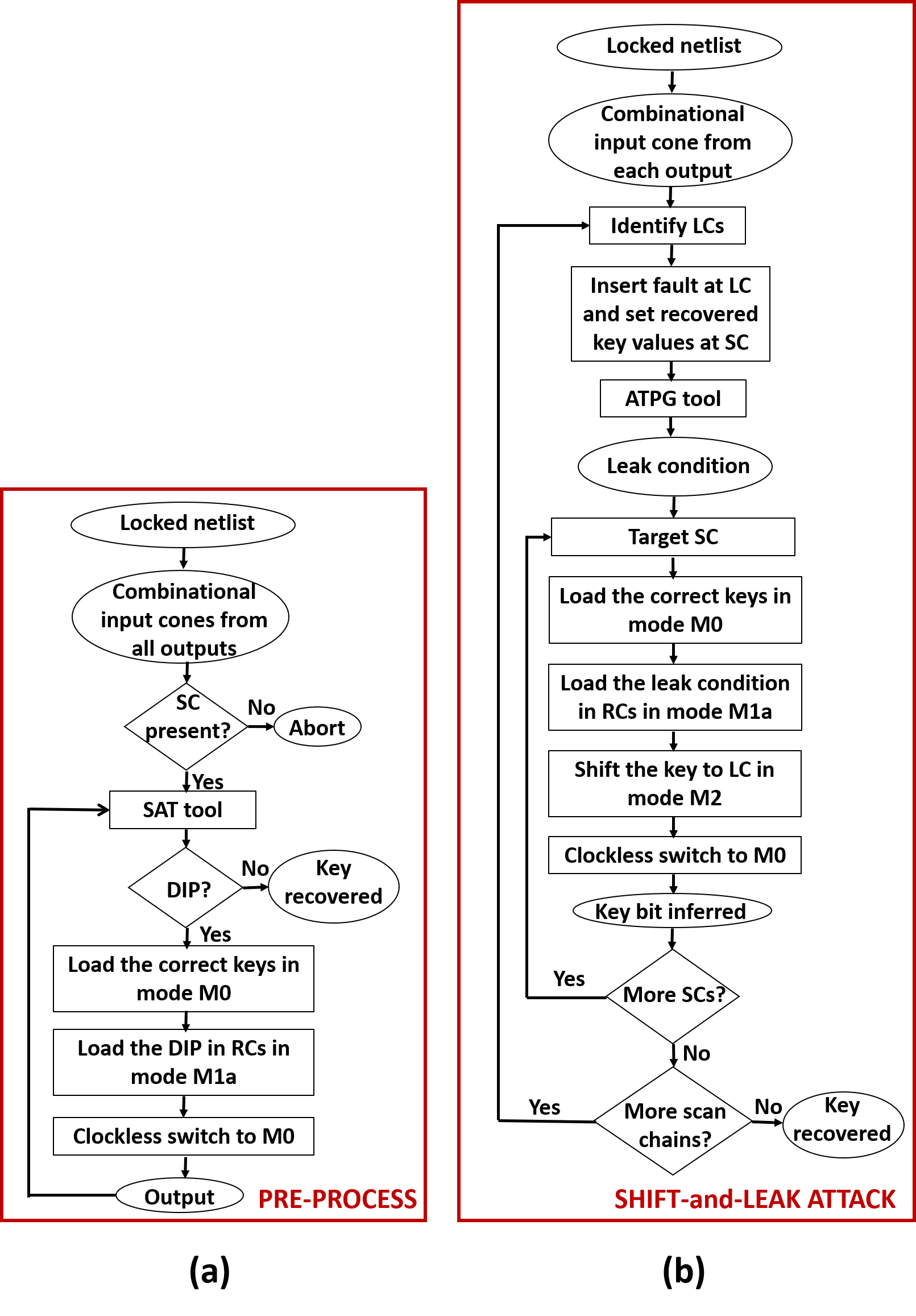}
    \caption{Flow chart for our methodology. (a) Pre-processing and (b) Shift-and-leak attack. 
    }
    \label{fig:flowchart}
\end{figure}

The methodology is shown in Fig~\ref{fig:flowchart}(b). We start from the LCs at the rightmost position in the scan chain as they can leak the highest number of SCs. This process is reiterated for all the scan chains in the design.

\begin{figure}[!tb]
    \centering
    \includegraphics[width=0.42\textwidth]{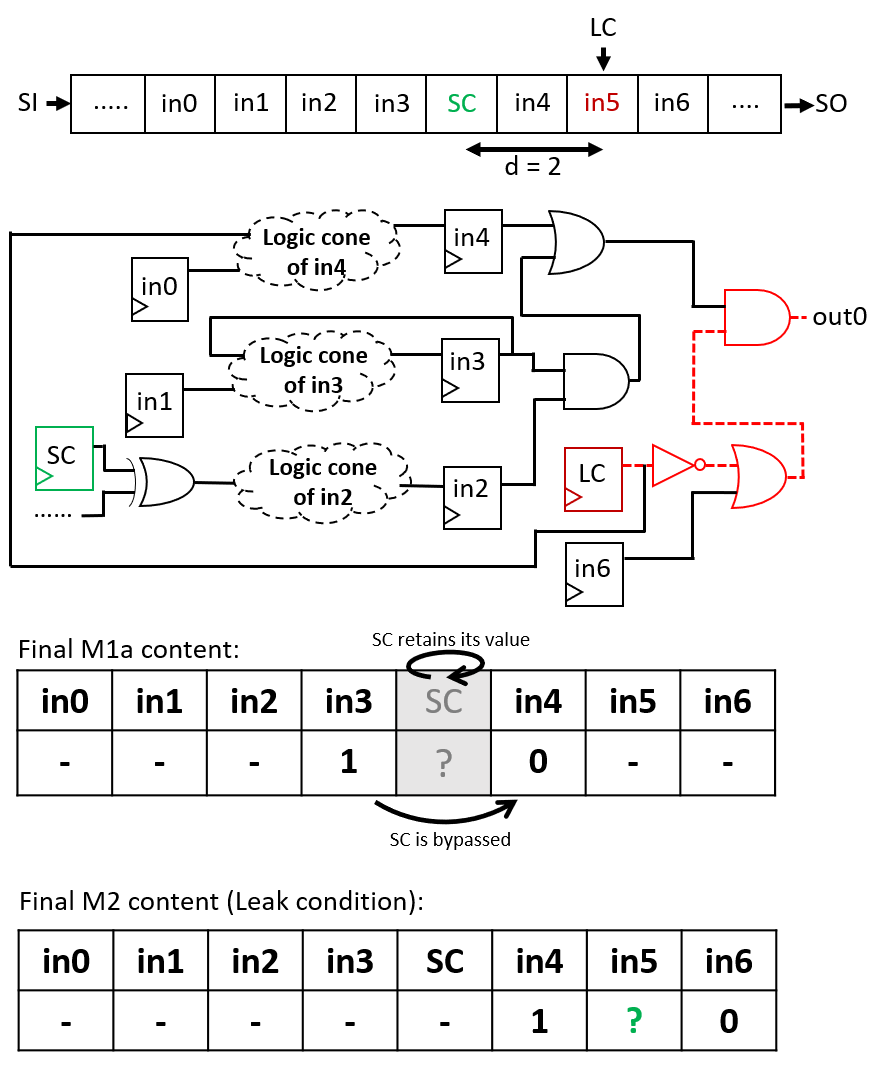}
    \caption{Example of a shift-and-leak attack on the DFS architecture. 
    $in5$ is LC, which can be propagated to the output $out0$ by setting two other bits in the scan chain (leak condition). The propagation path is marked in red and indicated by the dashed line. Post-$M1$a content required is a two-bit reverse-shifted version of the leak condition as $d=2$. In $M1$a, SC is not a part of the scan chain and thus, retains its content; in $M2$, SC becomes a part of the scan chain.}
    \label{fig:example_shiftandleak}
\end{figure}

\textcolor{black}{\textbf{Example.} Consider the example shown in Fig.~\ref{fig:example_shiftandleak}. The combinational fan-in cone of the primary output $out0$ includes five RCs ($in2$, $in3$, $in4$, $in5$, and $in6$), out of which we identify $in5$ as the LC. Using an ATPG tool, we insert a stuck-at-fault at the output of this LC and obtain the leak condition ($in4$ = 1 and $in6$ = 0) required to propagate this value to $out0$. From Fig.~\ref{fig:example_shiftandleak}, we observe that SC is of scan distance 2 to the LC; $d=2$. 
Next, we boot the chip in $M0$ and load the secret key from the secure memory into the SC. Then, we switch to $M1$a and load in the RCs the two-bit reverse-shifted version of the leak condition, while SCs hold their values; post $M1$a, $in3$ should be 1 and $in4$ should be 0. Now, we switch to $M2$ and perform a two-bit shift such that the leak condition reaches the correct RCs  \textcolor{black}{and SC reaches the LC}; $in4$ becomes 1, $in6$ becomes 0, \textcolor{black}{and $in5$ holds the key bit}. Finally, we make a clockless switch to M0 to leak the LC content to $out0$.}

The same LC can be re-used to retrieve the content of other SCs in the same scan chain to the left of LC; the process above is repeated but with a different $d$ value this time. 

\subsubsection{Limitations.} Identifying viable LCs may become challenging if the fan-in cones of the primary outputs include a large number of SCs. As they contain unknown key bits, the ATPG tool may fail to find a leak condition for a LC candidate in the presence of many SCs, limiting the number of viable LCs, and thus, the attack success as well. To circumvent this issue, we propose a pre-processing step that is described in the next section.

\subsection{Pre-process: Deciphering leaky secure cells}

\subsubsection{Idea.}
We can directly retrieve the content (key bits) of a group of SCs that are in the fan-in cone of a PO. This logic cone can be treated as a locked combinational circuit, on which SAT attack~\cite{Subramanyan_host_2015} can be applied without relying on scan-out reads. SAT attack can then produce distinguishing input patterns (DIPs) that need to be loaded \textcolor{black}{in} the scan chains in $M1$a mode; as the secret key held in SCs is overwritten with \textcolor{black}{the} stimulus in $M2$ mode, our attack rather relies on the shift operations in $M1$a mode. The working chip used as an oracle is only accessed on the PO under consideration; the scan-out ports are ignored. The response on the PO is used to eliminate incorrect keys that correspond to the group of SCs in the fan-in cone. DIP generation and key elimination are repeated until the correct key is found. 

\subsubsection{Methodology.} We follow a similar strategy to the previous attack where we exploit the primary output ports of the chip that are always observable. The attack is launched as follows.

\textcolor{black}{
\begin{enumerate}
    \item Extract the fan-in cones of \textcolor{black}{the} POs.
    \item Obtain the DIP from the SAT tool.
    \item Boot the IC in $M0$ mode, where SCs capture the secret key. 
    \item Switch to $M1$a to shift in the DIP obtained from the SAT tool into the RCs, while SCs retain their content, i.e., the secret key. 
    \item Clocklessly switch to $M0$ to obtain the response at the POs of the circuit. Feed the PO values to the SAT attack tool for key elimination, and go to step 2.
\end{enumerate}
}

The complete methodology is depicted in the flowchart in Fig.~\ref{fig:flowchart}(a). 

\begin{figure}[!tb]
    \centering
     \includegraphics[width=0.3\textwidth]{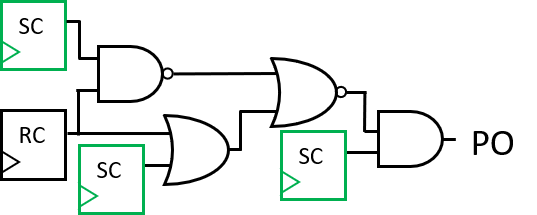}
    \caption{The fan-in cone of the PO consists of three SCs as three key inputs and one RC as a primary input. Identifying a leak condition for three unknown input values is challenging. Hence, the pre-processing step uses a SAT tool and tries to resolve these SCs prior to the shift-and-leak attack.
}
    \label{fig:SAT_example}
\end{figure}

\textcolor{black}{\textbf{Example.} Consider the combinational circuit controlled by three SCs and one RC, as shown in Fig.~\ref{fig:SAT_example}. These three SCs are part of the 128 SCs holding the complete key. Identifying a leak condition is challenging with three unknowns in a four-input circuit. However, we can obtain the values of these SCs using the SAT tool. Once these values are identified in the pre-processing step, we launch our shift-and-leak attack which successfully leaks the content of the remaining SCs as discussed earlier. }

Hence, the presence of SCs in the fan-in cone no longer poses any problem for the ATPG tool, as the content of these otherwise problematic SCs is already deciphered in the pre-processing phase.

\section{Proposed countermeasure} 

\subsection{Idea}
In our shift-and-leak attack, we exploit the $M1$a mode, \textcolor{black}{made} available in the DFS architecture. 
This mode was only used during the functional testing of the chip, but it was useful to shift-in \textcolor{black}{the} known patterns to apply our attack. 
Now, in order to thwart the proposed attack, it only makes sense to block the shift-in operation in $M1$a mode. 

With such a defense in place, RCs can no longer be controlled to desired (leak condition or DIP) values, effectively becoming no different than SCs that have unknown content. The proposed shift-and-leak attack that used to be able to resolve only SCs will then have to resolve both RCs and SCs. \textcolor{black}{Only the primary output fan-in cones that are mainly controlled by primary inputs and very few RCs and SCs will possibly reveal some information about the key, but that would be quite rare in typical designs.} We revisit this point quantitatively in Section~\ref{sec:exp_results}. 

\subsection{Architecture}
\begin{figure}[tb!]
    \centering
    \subfigure[]{\includegraphics[width=0.43\textwidth]{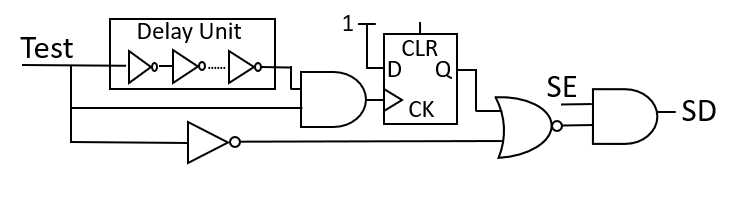}\label{fig:shift_control}}
    \subfigure[]{\includegraphics[width=0.3\textwidth]{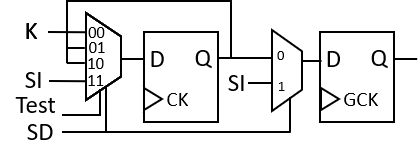}\label{fig:SE_shiftcontrol}}
    \caption{\textcolor{black}{Mode switch shift disable (MSSD)} countermeasure to prevent the proposed shift-and-leak attack. (a) When the Test pin is 0 or undergoes a positive transition, the output of the NOR gate goes low, and therefore, shift disable (SD) pin goes low. (b) SD pin instead of the SE pin controls the scan MUXes. During a positive transition on \textcolor{black}{the} Test pin, or when the Test pin is low, SD will become 0, thereby disabling \textcolor{black}{the} shift operation in the \textcolor{black}{RCs and SCs}.}
    \label{fig:defense}
\end{figure}

Figure~\ref{fig:defense} shows the new secure architecture, \textcolor{black}{mode switch shift disable (MSSD),} which blocks all the shift operations when Test pin is not asserted (Test = 0) or when there is a positive transition on the Test pin ($M0$ $\rightarrow$ $M1$b or $M0$ $\rightarrow$ $M2$). We assume 10 inverters in the delay unit, and the DFF sets to 0 on reset.\footnote{All the flip flops set to 0 on reset, unless otherwise mentioned.} After reset, when Test=1, the output of the single inverter will be 0. As the output of DFF will also be 0, the NOR output will go high, thereby, shift disable (SD) signal \textcolor{black}{will follow} the SE pin. The SC will now either be in $M1$b mode or in $M2$ mode.

In Table~\ref{tab:def}, we see NOR output and SD signal values for four cases where Test and SE can take on different values. When Test = 1, we observe that SD follows SE. This is correct only when the Test pin is high during power ON. However, when there is a positive transition on the Test pin or when the Test signal is low, NOR output goes low, and hence SD becomes 0, thereby restricting the shift operation. To conclude, we block the shift operation on both \textcolor{black}{the} SCs and RCs in $M1$a mode. There is no longer a mode where SCs can be bypassed, retaining their values, while RCs can be loaded/unloaded.

\begin{table}[!tb]
\caption{Shift disable (SD) values for different modes.}
\label{tab:def}
\centering
\begin{tabular}{|c|c|c|c|c|}
\hline
\textbf{Mode}   & \textbf{Test} & \textbf{SE} & \textbf{NOR} & \textbf{SD} \\ \hline\hline
M0              & 0             & 0           & 0            & 0                \\ \hline
M1a             & 0             & 1           & 0            & 0                \\ \hline
M1b             & 1             & 0           & 1            & 0                \\ \hline
M2              & 1             & 1           & 1            & 1                \\ \hline
\end{tabular}
\end{table}

\subsection{Test/debug operations}
$M1$a mode was used earlier for functional testing or in-field testing. We now have to analyze the impact of our \textcolor{black}{MSSD} defense on these tests.
For functional or in-field testing, we may need to bring the circuit to a known state and check its response. Earlier, this was made possible in mode $M1$a, where the SCs hold the correct key, and the test pattern is shifted in the RCs to bring the circuit to a known state. Now, by restricting the shift operation in $M1$a, we are creating hindrance in the functional testing of the chip. To solve this issue, we create a workaround \textcolor{black}{by} utilizing other modes. 

\begin{figure}[tb!]
    \centering
    \subfigure[]{\includegraphics[width=0.25\textwidth]{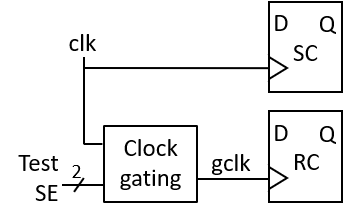}
    \label{fig:clock_gating_no_fsm}}
    \subfigure[]{\includegraphics[width=0.4\textwidth]{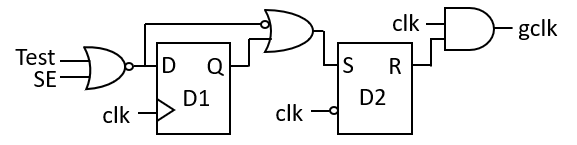}\label{fig:cg}}
    \subfigure[]{\includegraphics[width=0.35\textwidth]{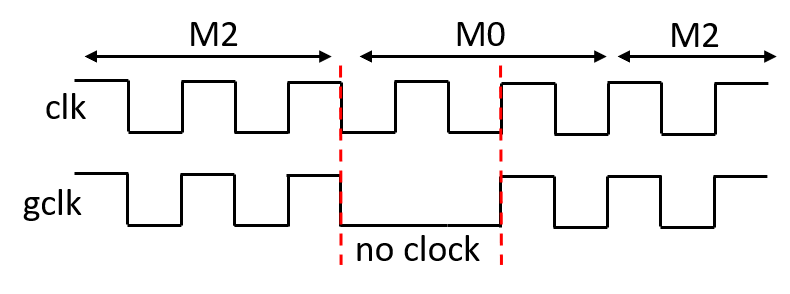}\label{fig:gated_clk}}
    \caption{(a) Architecture for gclk signal using clock gating (CG) circuitry. (b) The internal structure of the CG circuitry. D1 is a DFF which sets to 1 on reset, and D2 is a latch with a negative enable signal which also sets to 1 on reset. (c) Waveforms showing clk going to SCs and gclk going to RCs, along with \textcolor{black}{the} modes.}
    \label{fig:my_label}
\end{figure}

\textcolor{black}{As we no longer have the $M1$a mode, we rely on mode $M2$ to load the desired initial state into \textcolor{black}{the} RCs. The problem, however, is that the shift operations in $M2$ mode overwrites the SC content. A proper functional test requires that we bring the actual secret key from the tamper-proof memory into the SCs, which can only be done in mode $M0$, but that would overwrite the content of the RCs (initial state) as well. \textcolor{black}{We thus use clock gating in $M0$ to suppress the first clock pulse feeding the RCs;} this way, SCs are updated with the secret key while RCs maintain the desired initial state loaded in mode $M2$ prior to $M0$. Then in the next clock cycle, we resume the clock feeding these RCs which now operate in the presence of the correct key value. Figure~\ref{fig:clock_gating_no_fsm} explains how the clock is gated for one cycle for RCs, while the correct key is being loaded in the SCs. The clock gating (CG) circuit senses the switch from mode $M2$ to $M0$, and delays the clock feeding the RCs for one cycle; the clock feeding \textcolor{black}{the} SCs is untouched. Figure~\ref{fig:cg} shows the internal structure of the CG circuit and is explained in detail below. Figure~\ref{fig:gated_clk} shows the clock \textit{clk} and the gated clock \textit{gclk} signals along with \textcolor{black}{the} modes. }

\textbf{Clock gating. } After reset, both D1 and D2 are set to 1, therefore, the gclk signal follows the clk signal. 
\textcolor{black}{When the circuit is in $M2$ mode, the output of the NOR gate becomes 0, and hence, the output of the OR gate becomes 1. Thus, the output of D2 remains 1 and gclk signal follows clk signal. }
Now, when the mode is changed to $M0$, the output of the NOR gate becomes 1, therefore the output of the OR gate becomes 0 until D1 is updated. This OR output gets latched on to D2 \textcolor{black}{when the clock signal is low}. Hence, for one clock cycle, gclk goes low, as expected. After one clock cycle, the output of D1 is updated, and the OR output goes high again. Thus, gclk follows the clk signal once again. Further, when the circuit is booted in any mode, gclk follows the clk signal.

Although we block the shift operation on the SI pin in mode $M1$a, our defense, just like DFS, allows a debugger (without the knowledge of the correct key) to perform functional testing by observing the POs without any significant loss of coverage.\footnote{On average, we observed 0.03\% loss in test coverage w.r.t. the original design.} Further, if the debugger is trusted and knows the correct key, then s/he can directly load this key along with the pattern in mode $M2$ and capture its response in mode $M1$b, as in the DFS technique. 

\subsection{Security comparison} 
Upon unveiling and fixing the vulnerability of the robust DFS architecture, we now enable a scan access restriction based solution, \textcolor{black}{MSSD}, which when used with a very basic logic locking solution (say, Random Logic Locking (RLL) ~\cite{epic}), provides protection against all logic locking attacks. This would in turn deliver a much more comprehensive protection against not only scan attacks but also reverse engineering, IP piracy, Trojan insertion, etc. 

The techniques in~\cite{wang2019secure, ahlawat2017preventing} aim at protecting the scan access into \textcolor{black}{the} crypto cores based on a weaker threat model that assumes access to a working chip only; however, they are easily breakable in the threat model defined in this and all other logic locking papers where the locked netlist is also assumed to be available to an attacker. 

Prior logic locking defenses are either vulnerable to the SAT attack or fall short in delivering sufficient output corruption (e.g.,~\cite{sfll-fault,yasin_CCS_2017,anti-sat}), and thus are vulnerable to attacks such as AppSAT~\cite{appsat}.

Robust DFS in~\cite{guin2018robust} is vulnerable to our shift-and-leak attack; thus, it cannot protect the scan access of a logic-locked circuit. Our defense, on the contrary, disables both scan-in and scan-out access, preventing an attacker from launching any logic locking attack (e.g., SAT, AppSAT, etc.) that requires generating and applying input-output pairs by relying on scan access. 
 
\section{Experimental Results}\label{sec:exp_results}
\textcolor{black}{For our attack, }we analyzed eight different IWLS 2005 benchmarks~\cite{IWLS} \textcolor{black}{for five scan chain configurations,} locked via DFS using 128 key bits. Note that we apply strong logic locking~\cite{JV_DAC_2012} to lock these circuits as recommended by the authors~\cite{guin2018robust}. Further, for our proposed defense, \textcolor{black}{we implemented shift-blocking and CG circuitry from Figs.~\ref{fig:shift_control} and \ref{fig:cg} respectively,} on six benchmarks to obtain secure locked netlists.\textcolor{blue}{\footnote{We consider the same benchmarks mentioned in~\cite{guin2018robust} for uniform comparison. }} All the experiments have been carried out on a 128-core Intel Xeon processor running at 2.2 GHz with 256 GB of RAM, using Synopsys Design Compiler v16, Tetramax v16, and VCS v17 tools along with the 32nm SAED32 EDK Generic Library~\cite{32nm}. 

\subsection{Attack results}\label{sec:attack_results}
\subsubsection{Pre-processing}
Although the DFS technique claims to thwart the SAT attack by blocking scan read-out access, we still leverage the primary outputs to leak valuable information about the secret key. The attack results are shown in Table~\ref{tab:sat_sensi}. 
Note that as the success of pre-processing is dictated by the underlying structure of the locked circuit, it could only recover a few \textcolor{black}{key} bits for \texttt{s35932}, \texttt{s38417}, and \texttt{s38584}, and none for \texttt{b17}, \texttt{b18}, and \texttt{b19} \textcolor{black}{benchmarks.} \textcolor{black}{However, it could recover 47 and 120 key bits for \texttt{s13207} and \texttt{s15850}, respectively (irrespective of the scan chain configuration, i.e., the number of scan chains). For these two designs, the fan-in cones of the primary output ports contain a large number of SCs, i.e., 47 and 120, respectively, thereby allowing SAT solvers to directly access these key bits from the POs and decipher them. Further, test compressor and decompressor have no effect on our attack since we leverage primary outputs to leak the keys.}

\textbf{Run-time.} We present the run-time for pre-processing which consists of creating the combinational fan-in cones of the primary outputs of the circuit using the Synopsys Design Compiler tool and running the SAT solver. For all the cases, the analysis time is minimal, taking around one hundred seconds even for large designs ($>$10K gates) such as \texttt{s38584}.

\subsubsection{Shift-and-leak attack} 
After pre-processing, we launch our shift-and-leak attack, whose results are presented in Table~\ref{tab:sat_sensi} as well. On average, it recovers $\sim$95\% of the key-bits, sometimes reaching up to 100\%, completely undermining the security of the DFS architecture. The remaining key bits, if any, can be easily recovered via a brute-force attack.

\textbf{Effect of the number of scan chains.} Unlike the pre-processing step, the success of the shift-and-leak attack depends on the number of scan chains. For a larger number of scan chains, we see a monotonic decrease in \textcolor{black}{the} attack success. This can be attributed to the fact that the identified LCs can leak the content of a smaller number of SCs from their shorter scan chains. For example, there is a sharp drop in the number of deciphered key bits for \texttt{b17} with sixteen scan chains. 
\textcolor{black}{However, we observe that for \texttt{s15850} we recover the complete key even for sixteen scan chains. This is mainly because, we had deciphered most of the secure cells in the pre-processing step and those became the LCs to leak the rest of the key bits.}

\textbf{Run-time.} 
The run-time is dominated by the process of identifying LCs along with their leak conditions by executing the Tetramax tool. As can be seen from Table~\ref{tab:sat_sensi}, the attack takes only around 72 minutes even for the \texttt{b19} benchmark which has $>$100K gates.

\begin{table}[!tb]
\caption{Results of the \textcolor{black}{pre-processing step} and shift-and-leak attack launched on eight different IWLS-2005 benchmarks locked with a 128-bit key for different scan chain configurations: number of key bits recovered and analysis run-time.}
\label{tab:sat_sensi}
\setlength{\tabcolsep}{2pt}
\small\addtolength{\tabcolsep}{.0pt}
\centering
\begin{tabular}{|c|c|c|c|c|c|}
\hline
\multirow{2}{*}{\textbf{Benchmark}} & \multirow{2}{*}{\textbf{\begin{tabular}[c]{@{}c@{}}\#Scan \\ chains\end{tabular}}} & \multicolumn{2}{c|}{\textbf{Pre-process}}                                                 & \multicolumn{2}{c|}{\textbf{Shift-and-leak}}                                           \\ \cline{3-6} 
                                    &                                                                                    & \textbf{\begin{tabular}[c]{@{}c@{}}Key bits\\ recovered\end{tabular}}    & \textbf{\begin{tabular}[c]{@{}c@{}}Run-time\\ (secs)\end{tabular}} & \textbf{\begin{tabular}[c]{@{}c@{}}Key bits\\ recovered\end{tabular}} & \textbf{\begin{tabular}[c]{@{}c@{}}Run-time\\ (secs)\end{tabular}} \\ \hline\hline

\multirow{5}{*}{s13207}             & 1                                                                                  & \multirow{5}{*}{47}  & \multirow{5}{*}{28}                                                & 128               & \multirow{5}{*}{17}                                               \\ \cline{2-2} \cline{5-5}
                                    & 2                                                                                  &                      &                                                                    & 128               &                                                                    \\ \cline{2-2} \cline{5-5}
                                    & 4                                                                                  &                      &                                                                    & 128               &                                                                    \\ \cline{2-2} \cline{5-5}
                                    & 8                                                                                  &                      &                                                                    & 122               &                                                                    \\ \cline{2-2} \cline{5-5}
                                    & 16                                                                                 &                      &                                                                    & 101               &                                                                    \\ \hline
\multirow{5}{*}{s15850}             & 1                                                                                  & \multirow{5}{*}{120} & \multirow{5}{*}{36}                                                & 128               & \multirow{5}{*}{22}                                               \\ \cline{2-2} \cline{5-5}
                                    & 2                                                                                  &                      &                                                                    & 128               &                                                                    \\ \cline{2-2} \cline{5-5}
                                    & 4                                                                                  &                      &                                                                    & 128               &                                                                    \\ \cline{2-2} \cline{5-5}
                                    & 8                                                                                  &                      &                                                                    & 128               &                                                                    \\ \cline{2-2} \cline{5-5}
                                    & 16                                                                                 &                      &                                                                    & 128               &                                                                    \\ \hline
\multirow{5}{*}{s35932}             & 1                                                                                  & \multirow{5}{*}{3}   & \multirow{5}{*}{146}                                               & 127               & \multirow{5}{*}{124}                                               \\ \cline{2-2} \cline{5-5}
                                    & 2                                                                                  &                      &                                                                    & 127               &                                                                    \\ \cline{2-2} \cline{5-5}
                                    & 4                                                                                  &                      &                                                                    & 127               &                                                                    \\ \cline{2-2} \cline{5-5}
                                    & 8                                                                                  &                      &                                                                    & 127               &                                                                    \\ \cline{2-2} \cline{5-5}
                                    & 16                                                                                 &                      &                                                                    & 100               &                                                                    \\ \hline
\multirow{5}{*}{s38417}             & 1                                                                                  & \multirow{5}{*}{3}   & \multirow{5}{*}{120}                                               & 128               & \multirow{5}{*}{117}                                              \\ \cline{2-2} \cline{5-5}
                                    & 2                                                                                  &                      &                                                                    & 128               &                                                                    \\ \cline{2-2} \cline{5-5}
                                    & 4                                                                                  &                      &                                                                    & 128               &                                                                    \\ \cline{2-2} \cline{5-5}
                                    & 8                                                                                  &                      &                                                                    & 128               &                                                                    \\ \cline{2-2} \cline{5-5}
                                    & 16                                                                                 &                      &                                                                    & 80                &                                                                    \\ \hline
\multirow{5}{*}{s38584}             & 1                                                                                  & \multirow{5}{*}{8}   & \multirow{5}{*}{83}                                                & 128               & \multirow{5}{*}{87}                                              \\ \cline{2-2} \cline{5-5}
                                    & 2                                                                                  &                      &                                                                    & 128               &                                                                    \\ \cline{2-2} \cline{5-5}
                                    & 4                                                                                  &                      &                                                                    & 128               &                                                                    \\ \cline{2-2} \cline{5-5}
                                    & 8                                                                                  &                      &                                                                    & 128               &                                                                    \\ \cline{2-2} \cline{5-5}
                                    & 16                                                                                 &                      &                                                                    & 115               &                                                                    \\ \hline
\multirow{5}{*}{b17}                & 1                                                                                  & \multirow{5}{*}{0}   & \multirow{5}{*}{-}                                                 & 127               & \multirow{5}{*}{178}                                               \\ \cline{2-2} \cline{5-5}
                                    & 2                                                                                  &                      &                                                                    & 127               &                                                                    \\ \cline{2-2} \cline{5-5}
                                    & 4                                                                                  &                      &                                                                    & 127               &                                                                    \\ \cline{2-2} \cline{5-5}
                                    & 8                                                                                  &                      &                                                                    & 127               &                                                                    \\ \cline{2-2} \cline{5-5}
                                    & 16                                                                                 &                      &                                                                    & 36                &                                                                    \\ \hline
\multirow{5}{*}{b18}                & 1                                                                                  & \multirow{5}{*}{0}   & \multirow{5}{*}{-}                                                 & 126               & \multirow{5}{*}{1183}                                              \\ \cline{2-2} \cline{5-5}
                                    & 2                                                                                  &                      &                                                                    & 126               &                                                                    \\ \cline{2-2} \cline{5-5}
                                    & 4                                                                                  &                      &                                                                    & 126               &                                                                    \\ \cline{2-2} \cline{5-5}
                                    & 8                                                                                  &                      &                                                                    & 126               &                                                                    \\ \cline{2-2} \cline{5-5}
                                    & 16                                                                                 &                      &                                                                    & 126               &                                                                    \\ \hline
\multirow{5}{*}{b19}                & 1                                                                                  & \multirow{5}{*}{0}   & \multirow{5}{*}{-}                                                 & 127               & \multirow{5}{*}{4301}                                              \\ \cline{2-2} \cline{5-5}
                                    & 2                                                                                  &                      &                                                                    & 127               &                                                                    \\ \cline{2-2} \cline{5-5}
                                    & 4                                                                                  &                      &                                                                    & 127               &                                                                    \\ \cline{2-2} \cline{5-5}
                                    & 8                                                                                  &                      &                                                                    & 127               &                                                                    \\ \cline{2-2} \cline{5-5}
                                    & 16                                                                                 &                      &                                                                    & 127               &                                                                    \\ \hline
\end{tabular}
\end{table}
\subsubsection{Attack analysis}
\hspace*{\fill} \\
\textbf{Is our attack successful on DFS?} We are able to retrieve at least 80 out of 128 key bits for 97.5\% of the circuits. This essentially leaves the attacker to decipher $<$48 key bits, which can be simply achieved by applying a brute-force attack.\footnote{In today's computational standards, a key size of at least 80 bits is considered secure~\cite{security_key_size}.} Therefore, barring \texttt{b17} (for sixteen scan chains configuration), we can confidently consider most of the circuits (39 out of 40) broken. 

\textbf{Scalability.} Our attack terminates within a few minutes even for large designs such as \texttt{b19}, which contains $>$100K gates. Even for much larger circuits, we expect short attack run-times as ATPG is executed for a single fault per LC; to generate manufacturing test patterns for a chip, ATPG is executed for millions of faults! Moreover, the success rate of our shift-and-leak attack is \emph{independent of the design size}. As the identification of LCs and the associated leak conditions are linked to the testability of the design, we can expect a better success rate in highly testable designs.

\begin{figure}[!tb]
    \centering
    \includegraphics[width=0.4\textwidth]{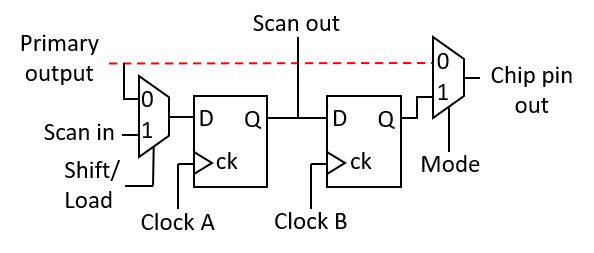}
    \vspace{-0.1in}
    \caption{Boundary scan cell design. Our attack utilizes the path marked in red when leaking the key value to a primary output in $M0$ mode. Source:~\cite{ieee1149}}
    \vspace{-0.1in}
    \label{fig:boundary_scan}
\end{figure}

\textbf{Attack in the presence of boundary scan.} The boundary scan technique is used to control and observe the primary output signal at the chip boundaries by placing shift-registers adjacent to the chip pins. The boundary scan cell design is shown in Fig.~\ref{fig:boundary_scan}. Note that the presence of boundary scan cells (which cannot be read-out) has no effect on the applicability of our attack. Our attack includes a clock-less switch into the functional ($M0$) mode, where the primary output must be observable on the chip pin to allow for functional debug; the signal path that enables our attack is illustrated (marked in red) in Fig.~\ref{fig:boundary_scan}.

\textbf{Comparison with prior attacks.}
Previously, there has been an attack launched on camouflaged sequential circuits with restricted scan access~\cite{massad17}.
To circumvent the blocked scan access, El Massad et al. rely on model checker tools to find \emph{discriminating input sequences} that can be applied through multiple capture cycles to observe the primary output. However, their attack runs into scalability issues as it relies on two sub-routines which are in PSPACE and NP. As a result, their attack fails to terminate for fairly small designs such as \texttt{s5378} and \texttt{s9234}, which contain only a few thousand gates. 
\textcolor{black}{Recently, a faster sequential deobfuscation approach was presented, which reduces the run-time of sequential SAT attacks by two orders of magnitude~\cite{kc2}. However, their approach takes around two hours to deobfuscate the \texttt{s35932} benchmark; our attack takes only 10 minutes to leak 127 key bits.} Table~\ref{tab:compare_priorwork} shows the attack success rate and average time taken by~\cite{massad17},~\cite{kc2}, and our shift-and-leak attack for various circuits~\cite{IWLS}. Further, our attack takes lesser time to break larger circuits than what previous techniques take to break smaller circuits.
While the scalability of the prior approaches is questionable, our attack can be applied in a straight-forward manner by running the ATPG tool to generate a single pattern (leak condition) per LC without any scalability issues.

\begin{table}[!tb]
\caption{Comparison between~\cite{massad17},~\cite{kc2}, and our shift-and-leak attack in terms of time taken (in seconds) to recover the whole key. Time taken for our attack includes the pre-processing step as well. Results for the larger circuits were not provided by the previous techniques, hinting at scalability issues.}
\label{tab:compare_priorwork}
\setlength{\tabcolsep}{1pt}
\small\addtolength{\tabcolsep}{.0pt}
\begin{tabular}{|c|c|c|c|c|c|}
\hline
\multirow{2}{*}{\textbf{Benchmark}} & \multirow{2}{*}{\textbf{\begin{tabular}[c]{@{}c@{}}Capacity\\ (\# Gates)\end{tabular}}} & \multicolumn{4}{c|}{\textbf{Time taken (secs)}} \\ \cline{3-6} 
                           &                           & \textbf{NuSMV~\cite{massad17}}      & \textbf{nuXmv~\cite{kc2}}      & \textbf{Int~\cite{kc2}}      & \textbf{Proposed}     \\ \hline\hline
s15850                     & 9772                      & -                          & -                     & 54                  & 58           \\ \hline
s35932                     & 16065                     & -                          & 8268                  & 3885                & 270          \\ \hline
b17                        & 32326                     & -                          & -                     & -                   & 178          \\ \hline
b18                        & 114621                    & -                          & -                     & -                   & 1183          \\ \hline
b19                        & 231320                    & -                          & -                     & -                   & 4301          \\ \hline
\end{tabular}
\end{table}

\subsection{Defense results}

\begin{table*}[!tb]
\caption{Area overhead along with test coverage comparison between original, DFS, and MSSD designs for identical timing constraints. Number of key bits (out of 128) recovered in DFS and MSSD architectures are presented for a single scan chain (best case scenario for the attack). \textcolor{black}{We re-synthesized the DFS designs in our setup for uniform comparison.}}
\label{tab:atf}
\setlength{\tabcolsep}{3pt}
\centering
\small\addtolength{\tabcolsep}{.0pt}
\begin{tabular}{|c|c|c|c|c|c|c|c|}
\hline
\multirow{2}{*}{\textbf{Benchmark}} & \textbf{Original}      & \multicolumn{3}{c|}{\textbf{DFS}}                                                                                       & \multicolumn{3}{c|}{\textbf{Proposed (MSSD)}}                                                                                  \\ \cline{2-8} 
                                    & \textbf{\begin{tabular}[c]{@{}c@{}}Test coverage\\ (\%)\end{tabular}} & \textbf{\begin{tabular}[c]{@{}c@{}}Area overhead\\ (\%)\end{tabular}} & \textbf{\begin{tabular}[c]{@{}c@{}}Test coverage\\ (\%)\end{tabular}} & \textbf{\begin{tabular}[c]{@{}c@{}}Key bits\\ recovered\end{tabular}} & \textbf{\begin{tabular}[c]{@{}c@{}}Area overhead\\ (\%)\end{tabular}} & \textbf{\begin{tabular}[c]{@{}c@{}}Test coverage\\ (\%)\end{tabular}} & \textbf{\begin{tabular}[c]{@{}c@{}}Key bits\\ recovered\end{tabular}} \\ \hline\hline
s35932                              & 100                    & 6.98                   & 100                    & 127                                                                   & 3.01                   & 100                    & 0                                                                     \\ \hline
s38417                              & 100                    & 7.73                   & 100                    & 128                                                                   & 3.74                   & 100                    & 0                                                                     \\ \hline
s38584                              & 100                    & 9.07                   & 100                    & 128                                                                   & 5.48                   & 100                    & 0                                                                     \\ \hline
b17                                 & 99.91                  & 4.78                   & 99.72                  & 127                                                                   & 1.49                   & 99.69                  & 0                                                                     \\ \hline
b18                                 & 99.77                  & 1.38                   & 99.78                  & 126                                                                   & -1.34                  & 99.77                  & 0                                                                     \\ \hline
b19                                 & 99.8                  & 0.97                   & 99.78                  & 127                                                                   & -2.51                  & 99.79                  & 0                                                                     \\ \hline
\end{tabular}
\end{table*}

\textcolor{black}{Our} proposed architecture consists of shift-blocking and CG circuitry to thwart the shift-and-leak attack and yet enables functional testing. \textcolor{black}{We computed area overhead for DFS and MSSD with respect to original designs, for identical timing constraints.}
We also computed test coverage for the original, DFS, and proposed designs.

\textcolor{black}{\textbf{Security analysis.}
When \textcolor{black}{our MSSD} defense is applied with \textit{RLL}, we observe that no key bits are recovered when the shift-and-leak attack is applied on it, as shown in Table~\ref{tab:atf} column 8. As we restrict the shifting of scan patterns in $M1$a, the shift-and-leak attack recovers no key bits, irrespective of benchmarks and scan configurations. 
As expected, there are no PO fan-in cones in these benchmark circuits that are fed by mainly PIs and only a few RCs and SCs, resulting in no viable LCs.}

\textbf{Area overhead.}
Area overhead for six benchmarks for DFS and \textcolor{black}{MSSD} is presented \textcolor{black}{in Table~\ref{tab:atf}, columns 3 and 6.} Most of the area overhead comes from the secure cell components. On average, we obtain $\sim$5.15\% area overhead for the DFS technique and $\sim$1.65\% for our proposed technique. \textcolor{black}{As observed, our proposed technique has lower area overhead than DFS; our method implements only one instance of CG circuitry in place of a series of scan read blocking OR gates that the DFS architecture uses.}
Further, we observe that our proposed design has a lower area than the original design for benchmarks \texttt{b18} and \texttt{b19}. This is because the Synopsys Design Compiler tool optimizes the circuit with the added gates.

\textbf{Test coverage.}
Test coverage is calculated when the chip is in test mode, and hence the clock signal is implicitly tested. On average, we observe that test coverage for DFS and our proposed architectures are almost the same ($\sim$99.88\%) and only 0.03\% less than that \textcolor{black}{of} the original design. 

\section{Conclusion}
A cost-effective robust DFS technique presented in~\cite{guin2018robust} aims at thwarting all \textcolor{black}{existing} logic locking attacks, including the powerful SAT attack, by restricting the scan access. 

In this work, we propose an attack that exploits the offered capabilities in leaking the key bits despite the restricted scan access. Our technique first identifies leaky cells and then uses the shift-and-leak attack to leak the key values to primary output ports. 
\textcolor{black}{We demonstrated our attack on eight different IWLS-2005 benchmarks for five scan chain configurations.} On average $\sim$95\% of the key bits are leaked, and 100\% in most cases. To thwart this attack, we propose a defense, which maintains the same testability as \textcolor{black}{the} DFS architecture but with lower area, while delivering resilience against all other attacks. \textcolor{black}{We compared our attack and defense with prior works and showed how we significantly advance the state of the art through our novel attack and defense.} A comprehensive set of attacks ranging from scan attacks to reverse engineering, IP piracy, and Trojan insertion can all be thwarted \textcolor{black}{by our defense.} 
 
\bibliographystyle{IEEEtran}
\bibliography{main.bib}

\end{document}